%% file: manuscript.tex
\documentclass[aps,prl,reprint,superscriptaddress,notitlepage,nobibnotes]{revtex4-2}

\usepackage{amsmath}
\usepackage{amssymb}
\usepackage{graphicx}
\usepackage{xcolor}
\newcommand{\revision}[1]{#1}
\usepackage{microtype}
\usepackage[hidelinks]{hyperref}

\graphicspath{{figures/}}

\begin{document}

\title{Laser-Driven Electron Emission Carrying Orbital Angular Momentum from Carbon Nanotubes}

\author{Lihan Chi}
\affiliation{School of Physics, Hefei University of Technology, Hefei, China}
\affiliation{School of Physics, Nanjing University of Science and Technology, Nanjing, China}

\author{Ziwen Wang}
\affiliation{Department of Physics, The University of Hong Kong, Hong Kong, China}

\author{Yigeng Peng}
\email{pengyg@hfut.edu.cn}
\affiliation{School of Physics, Hefei University of Technology, Hefei, China}

\author{Chao Yu}
\email{chaoyu@njust.edu.cn}
\affiliation{School of Physics, Nanjing University of Science and Technology, Nanjing, China}

\author{Zhongjun Li}
\affiliation{School of Physics, Hefei University of Technology, Hefei, China}

\author{Ruifeng Lu}
\email{rflu@njust.edu.cn}
\affiliation{School of Physics, Nanjing University of Science and Technology, Nanjing, China}

\input{sections/abstract}

\maketitle

\input{sections/introduction}

\input{sections/methods}

\input{sections/results}

\input{sections/conclusion}
\input{sections/acknowledgments}

\bibliography{references}

\end{document}


\title{Supplemental Material for\\``Laser-Driven Electron Emission Carrying Orbital Angular Momentum from Carbon Nanotubes''}
\author{Lihan Chi}
\affiliation{School of Physics, Hefei University of Technology, Hefei, China}
\affiliation{School of Physics, Nanjing University of Science and Technology, Nanjing, China}
\author{Ziwen Wang}
\affiliation{Department of Physics, The University of Hong Kong, Hong Kong, China}
\author{Yigeng Peng}
\email{pengyg@hfut.edu.cn}
\affiliation{School of Physics, Hefei University of Technology, Hefei, China}
\author{Chao Yu}
\email{chaoyu@njust.edu.cn}
\affiliation{School of Physics, Nanjing University of Science and Technology, Nanjing, China}
\author{Zhongjun Li}
\affiliation{School of Physics, Hefei University of Technology, Hefei, China}
\author{Ruifeng Lu}
\email{rflu@njust.edu.cn}
\affiliation{School of Physics, Nanjing University of Science and Technology, Nanjing, China}
\maketitle
\section{Computational parameters}

The carbon nanotube (CNT) geometries were hydrogen-terminated armchair nanotubes corresponding to CNT(3,3) and CNT(5,5), with compositions C$_{78}$H$_{12}$ and C$_{130}$H$_{20}$, respectively. These models contain 324 and 540 valence electrons. The hydrogen termination and finite length provide a well-defined finite-cell representation and are not assumed to be essential ingredients of the helicity-to-OAM conversion mechanism. The oriented CO$_2$ reference contains 16 valence electrons. The benzene reference contains 30 valence electrons and is placed in the $yz$ plane, so that the molecular normal and the OAM analysis axis are both $x$.

The CNT calculations used $50\times30\times30$~\AA$^3$ simulation boxes with the tube axis aligned along $x$. The CO$_2$ reference used a $20\times16\times16$~\AA$^3$ box with the molecular axis aligned along the same direction, and benzene used a $20\times20\times20$~\AA$^3$ box. The real-space grid spacing was 0.3 a.u., and the electron-ion interaction was described with Hartwigsen-Goedecker-Hutter (HGH) pseudo-potentials~\cite{Hartwigsen1998HGH} within the local density approximation (LDA)~\cite{Onida2002Excitations}. Time propagation used the enforced time-reversal-symmetry propagator with Lanczos exponentiation, a time step of 0.05 a.u., and output every 50 steps. Symmetry constraints were disabled during propagation. Outgoing density was absorbed by an $x$-directed complex absorbing potential (CAP) with height of $-2.5$ a.u. and a linear $|x|$ profile spanning from 20.0 to 25.0~\AA~\cite{DeGiovannini2015Absorbing}.

\begin{figure}[htbp]
\centering
\includegraphics[width=0.65\linewidth]{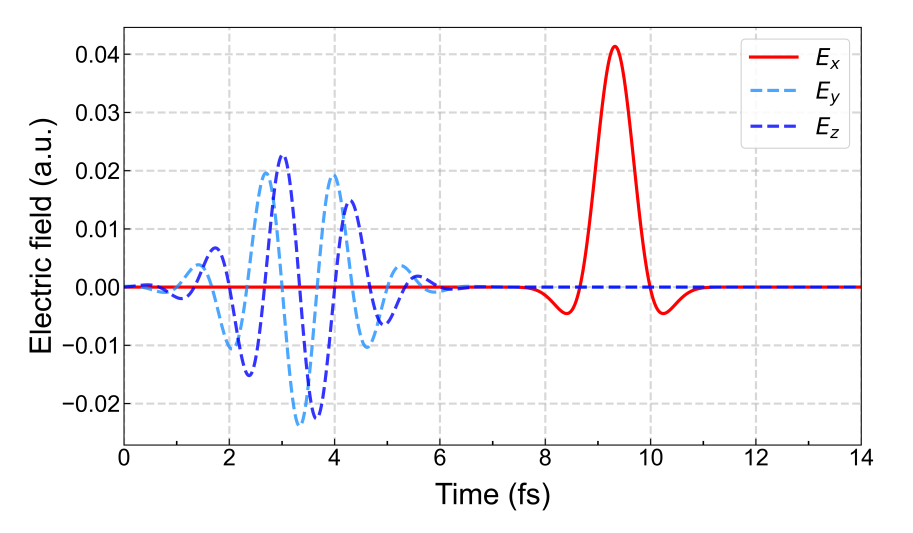}
\caption{Temporal profile of the laser electric field components. The OAM-writing pulse (blue dashed lines) is circularly polarized in the $yz$ plane. The extraction pulse (red solid line) is linearly polarized along the $x$-axis. The two pulses are separated by a time delay of 6.0 fs.}
\label{fig:s1}
\end{figure}

The laser field consisted of a circular OAM-writing pulse followed by a delayed linear extraction pulse. The OAM-writing pulse was a 400 nm pulse circularly polarized in the $yz$ plane, with full width at half maximum (FWHM) 2.4 fs, peak intensity $2.0\times10^{13}$ W cm$^{-2}$, and field maximum at 3.32 fs. The extraction pulse was an 800 nm pulse linearly polarized along $x$, with FWHM 1.2 fs, peak intensity $6.0\times10^{13}$ W cm$^{-2}$, and field maximum at 9.32 fs, giving a 6.0 fs pump-extraction delay (see Fig.~\ref{fig:s1}). The carrier-envelope phase of the extraction pulse was chosen so that the emitted density is driven predominantly toward one side of the simulation box.

\section{Angular-momentum and integration region definitions}

The axial orbital angular momentum is measured about the tube axis. The relation
\[
L_x=y p_z-z p_y
\]
fixes the axis and sign convention. For the time-dependent simulations with nonlocal HGH pseudopotentials, we did not evaluate a bare canonical local $\mathbf{r}\times\mathbf{p}$ density. Instead, the OAM output follows the magnetic-response pathway used by OCTOPUS for the total time-dependent angular momentum: a velocity-form orbital angular momentum operator with the default gauge-including projector augmented-wave (GIPAW) correction~\cite{Pickard2001MagneticResponse}. The unmodified OCTOPUS \revision{\texttt{angular}} output gives only the total $L_x,L_y,L_z$. We therefore used a local extension that exposes the same operator's grid-resolved contribution and accumulates it over Cartesian slices,
\[
L_x^{\mathrm{line}}(x,t)=\int l_x(x,y,z,t)\,dy\,dz,
\]
where $l_x$ is the axial-OAM contribution from the magnetic-response evaluation. Summing all slices recovers the corresponding total $L_x$ to within grid-discretization error, so the total angular output, the time-dependent angular-momentum-distribution (TDAMD), $\Lcol(t)$, and $\Lmean(t)$ use a single numerical definition of OAM. The spatially resolved TDAMD is used as an operational partition of this angular-momentum expectation over the simulation cell.

\revision{The external integration region lies beyond the rightmost atom of each system,}
\[
\revision{x>x_{\max}+2.5\,\text{\AA},}
\]
\revision{where $x_{\max}$ is the $x$-coordinate of that atom.} The emitted electron number in this region is
\[
n(t)=\int_{\revision{x>x_{\max}+2.5\,\text{\AA}}}\rho(\mathbf{r},t)\,d\mathbf{r},
\]
with $\rho(\mathbf{r},t)$ normalized to the valence-electron count of the finite system. The ground-state density in this region is negligible on the scale used here, so $n(t)$ is equivalent to the excess emitted electron number on the plotted scale. Applying the same spatial cut to the TDAMD defines the collected axial OAM and the mean axial OAM carried per ionized electron in the integration region,
\[
\Lcol(t)=\int_{\revision{x>x_{\max}+2.5\,\text{\AA}}}L_x^{\mathrm{line}}(x,t)\,dx,
\]
\[
\Lmean(t)=\Lcol(t)/n(t).
\]
$\Lcol(t)$ is reported in $\hbar$, and $\Lmean(t)$ in $\hbar/e$.
\section{Circular-pump-removed control}

\begin{figure}[htbp]
\centering
\includegraphics[width=0.70\linewidth]{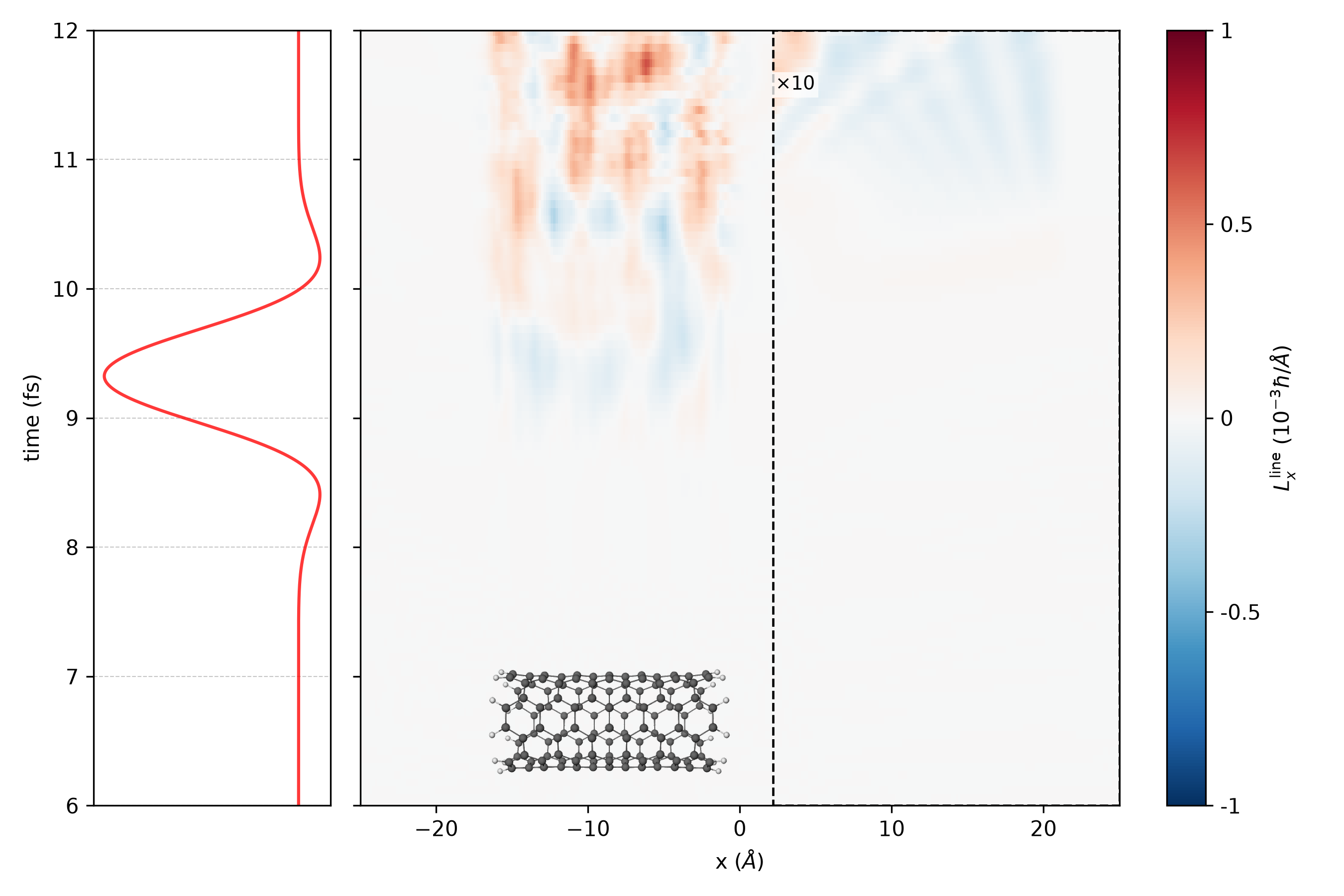}
\caption{TDAMD for the CNT(5,5) circular-pump-removed control. The linearly polarized extraction pulse is retained, but the circular preparation field is absent. \revision{The red curve on the left shows the extraction-pulse profile. The TDAMD is shown over 6--12 fs, with a color scale of $\pm10^{-3}\hbar/\text{\AA}$. The dashed box marks the external integration region, $x>x_{\max}+2.5\,\text{\AA}$, where the TDAMD is multiplied by 10 for visibility. The weak response is localized near the CNT; no resolved OAM-carrying emission feature appears in the external integration region.}}
\label{fig:s2}
\end{figure}

As a control for the external integration region analysis window, we performed calculations with the circular preparation step removed while retaining the linearly polarized extraction pulse. \revision{The expanded TDAMD color scale in Fig.~\ref{fig:s2} reveals a weak response near the CNT, but no resolved OAM-carrying emission feature in the external integration region.} Consequently, the collected electron number is too small for a meaningful emitted-electron-normalized $\Lmean=\Lcol/n$ comparison.

For this reason the circular-pump-removed control is not included as a main-text panel. Its role is limited to checking that the reported signal in the external integration region is not an artifact of the analysis window or the ground-state density tail. The decisive maintext symmetry test is instead the helicity reversal in CNT(5,5), where the emitted charge is essentially unchanged while $\Lcol$ and $\Lmean$ change sign.

\section{Axial kinetic-energy estimate}

\begin{figure}[htbp]
\centering
\includegraphics[width=0.94\linewidth]{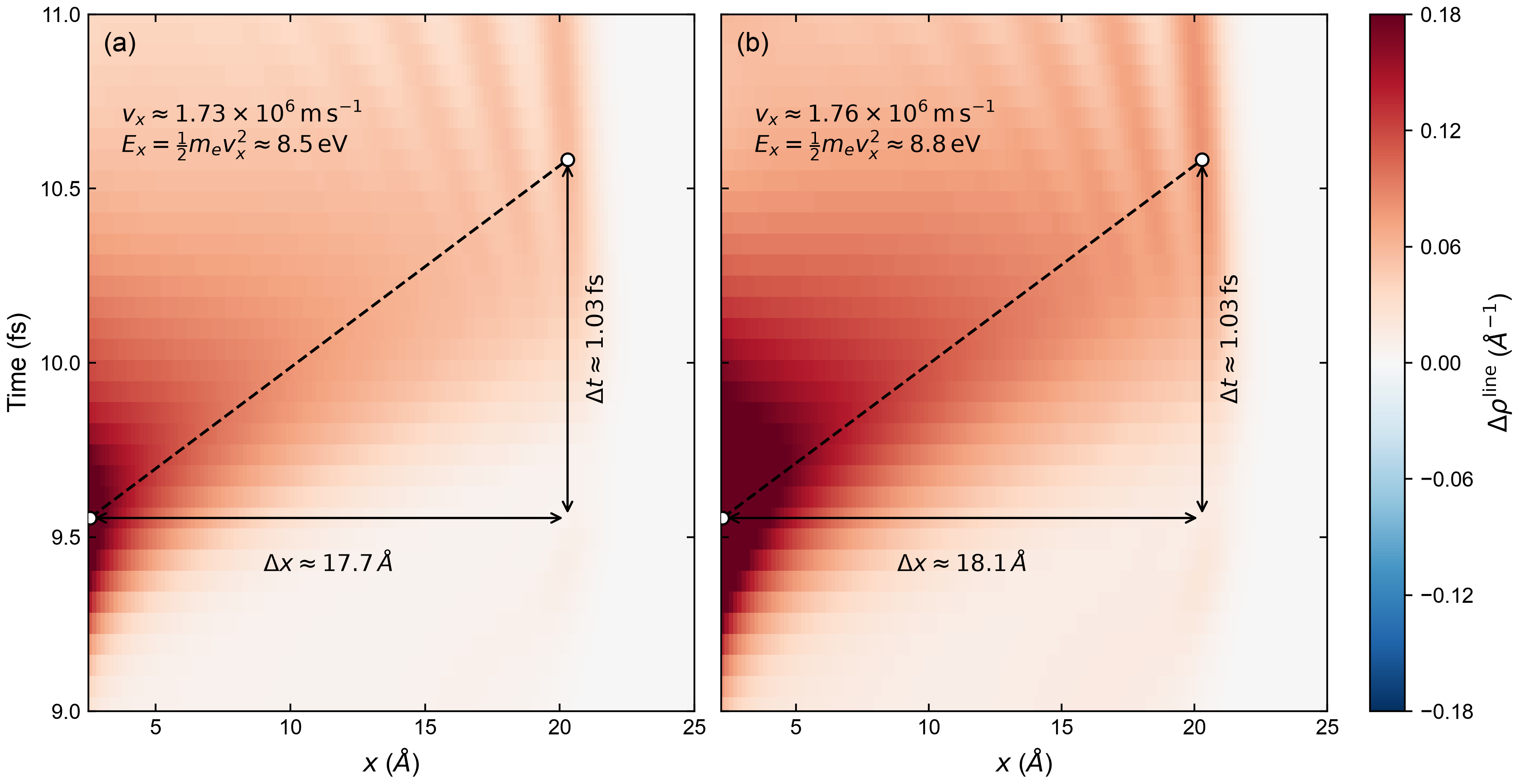}
\caption{Time-resolved density-line maps of the emitted electron wave packet for (a) CNT(3,3) and (b) CNT(5,5). The black dashed line, with open circles at its endpoints, traces the trajectory of the emitted-density peak. The black solid lines with arrowheads mark the axial displacement $\Delta x$ and time interval $\Delta t\approx1.03$ fs, from which the axial electron velocity $v_x$ and the corresponding free-electron axial kinetic energy $E_x$ are estimated. The color bar gives differential electron line density $\Delta\rho^{\mathrm{line}}$ in \AA$^{-1}$.}
\label{fig:s3}
\end{figure}

The main text describes the emitted CNT wave packet as a slow electron wave packet. Instead of computing the energy-resolved photoelectron spectrum, as a conservative scale estimate, we read the propagation speed from the trajectory of the emitted-density peak in the density-line maps. The marked displacements in Fig.~\ref{fig:s3} are $\Delta x\approx17.7$~\AA{} for CNT(3,3) and $\Delta x\approx18.1$~\AA{} for CNT(5,5), both over $\Delta t\approx1.03$ fs, corresponding to $v_x\approx1.73\times10^6$~\mbox{m s$^{-1}$} and $v_x\approx1.76\times10^6$~\mbox{m s$^{-1}$}, respectively.

Using a free-electron axial kinetic-energy estimate,
\[
E_x\approx\frac{1}{2}m_e v_x^2,
\]
this gives $E_x\approx8.5$ eV for CNT(3,3) and $E_x\approx8.8$ eV for CNT(5,5). We therefore use ``slow electron'' in the main text in the sense of emitted electron wave packets with energies of a few to ten eV.

\bibliography{supplemental-references}

%% file: sections/abstract.tex
\begin{abstract}

Extending the orbital angular momentum (OAM) degree of freedom to slow electrons would open a distinct regime of low-energy electron--matter interactions. Using time-dependent density-functional theory, we demonstrate that laser-driven armchair carbon nanotubes (CNTs) can emit \revision{OAM-carrying slow electrons}. A circularly polarized OAM-writing pulse promotes circumferential electronic motion and generates OAM about the tube axis; a delayed linearly polarized extraction pulse releases slow electrons carrying the injected angular momentum. \revision{We find much higher ionization yields and net emitted axial OAM in representative CNTs than in oriented CO\(_2\) and benzene under identical laser conditions. Increasing the tube diameter enhances the ionization yield and OAM substantially, and moreover, reversing the helicity of the OAM-writing pulse flips the sign of the emitted OAM.} Together, these findings establish CNTs as a platform for slow-electron emission with optically controllable OAM.

\end{abstract}

%% file: sections/introduction.tex
Electron vortex beams are structured free-electron waves that carry orbital angular momentum (OAM) about their propagation axis~\cite{Bliokh2017FreeElectronVortex,Lloyd2017ElectronVortices}. Conventional electron vortex beams have primarily been produced in electron microscopes at kinetic energies of tens to hundreds of kiloelectronvolts, using electron-optical elements such as spiral phase plates and holographic masks to imprint a helical phase~\cite{Uchida2010Nature,Verbeeck2010Nature,McMorran2011Science}. These high-energy vortex electrons have been used in magnetic spectroscopy, phase-contrast imaging, and angular-momentum-transfer studies~\cite{Bliokh2017FreeElectronVortex,Lloyd2017ElectronVortices,Lloyd2012MagneticDichroism,AsenjoGarcia2014Dichroism}. By contrast, slow electrons in the electronvolt-to-tens-of-electronvolts range interact with matter differently and play central roles in resonant molecular scattering, dissociative electron attachment, surface charge transfer, and radiation-damage chemistry~\cite{Gorfinkiel2017,Fabrikant2016}. Extending OAM to the slow-electron regime therefore requires more than lowering the energy of conventional electron vortex beams. It introduces OAM as a controllable degree of freedom into a distinct class of low-energy electron--matter interactions.

For established keV electron beams, OAM generation and control have largely relied on external shaping or coupling after beam formation. In addition to spiral phase plates and holographic masks, approaches based on membranes, plasmonic near fields, surface modes, and laser--electron scattering provide active control over free-electron wavefronts, OAM, and spatiotemporal structure~\cite{Vanacore2019,Tsesses2023PhotonInducedModulation,Fang2026,Bu2024}. These methods, however, are difficult to apply directly to slow electrons. Once emitted, slow electrons are highly susceptible to interfaces, stray fields, surface charging, and near-field perturbations, which makes post-emission manipulation particularly challenging. These constraints motivate a different strategy: preparing electron OAM about a defined axis inside the emitter before ionization, then releasing the electrons as OAM-carrying slow electrons.

Light--matter interactions offer a natural route toward this goal. A bound target provides the coupling environment needed to transfer angular momentum from the driving field to the electronic system before or during electron release. Related optically driven routes to OAM-carrying slow electrons have been explored in atomic and molecular systems. In photoionization by twisted light, the light field can transfer its OAM to photoelectrons; for targets on the beam axis and ideally oriented, the outgoing electron can carry a well-defined OAM projection~\revision{\cite{Pavlov2024AtomicPhotoionizationVortex}}. Strong-field studies have also predicted the generation of \revision{OAM-carrying slow electrons} through above-threshold ionization in circularly polarized laser fields~\cite{Kang2021}. Tunneling ionization has likewise been proposed to release \revision{OAM-carrying slow electrons} under specific target conditions, including initial OAM-biased orbitals and multicenter molecular geometries~\cite{Bazarov2023,Bazarov2024}. \revision{Related theoretical work has extended vortex-electron concepts to rescattering spectroscopy and target-structure imaging~\cite{Tolstikhin2019}, ultrafast chiral imaging~\cite{Planas2022StrongFieldChiralImaging}, and OAM-resolved molecular photoionization~\cite{Bazarov2025}.} More broadly, attoclock interferometry can reconstruct the spatiotemporal amplitude and phase of photoelectron wave functions~\cite{Ge2024AttoclockInterferometry}, providing a complementary diagnostic framework for future OAM-resolved studies. Attosecond vortex photoelectron holography has further been proposed to retrieve phase-sensitive chiral information from OAM-carrying photoelectrons~\cite{Li2026VortexPhotoelectronHolography}. Beyond atomic and molecular ionization, surface photoemission driven by twisted photons has also been proposed as a material-based route to generating OAM-carrying electrons~\cite{Kazinski2025}.

Direct generation of OAM-carrying slow electrons through light--matter interactions imposes specific requirements on the emitter. The emitter should provide a well-defined axis and sufficiently high rotational symmetry for axial OAM to remain a good quantum number, or at least to suppress mixing among OAM components. Atoms and linear molecules provide the highest axial symmetry. However, their limited electronic extent perpendicular to the symmetry axis restricts the transition-dipole matrix elements for circumferential motion and thus weakens coupling to circularly polarized light. Their discrete electronic spectra also make optical pumping sensitive to the laser frequency and specific resonances. Cyclic systems with delocalized \(\pi\) electrons offer a more practical compromise. Their discrete rotational symmetry may introduce some mixing among OAM components, but their larger transverse extent and delocalized electronic structure can support stronger transition-dipole coupling to circumferential motion. Although finite cyclic systems generally retain a highest occupied molecular orbital--lowest unoccupied molecular orbital (HOMO--LUMO) gap that constrains the pump frequency, their denser manifold of excited states provides more optically accessible pumping channels than small, highly symmetric systems. Efficient OAM injection alone is insufficient; the prepared electrons must also be released without strongly perturbing their angular-momentum distribution. An emitter that is difficult to ionize requires a stronger extraction field. Such a field can generate background electrons without prepared OAM and open ionization channels carrying axial OAM of the opposite sign, thereby weakening the predominance of one OAM sign and reducing the net emitted OAM. \revision{A suitable source of OAM-carrying slow electrons} must therefore balance axial symmetry, circumferential optical coupling, access to excited states, and efficient electron release.

In this context, carbon nanotubes offer a promising material platform for \revision{emission of OAM-carrying slow electrons}. Their cylindrical geometry defines a material axis, while their rotational symmetries provide quantum numbers for circumferential electronic motion~\cite{Damnjanovic1999}, thereby enabling efficient coupling of circularly polarized laser fields to such motion~\cite{Samsonidze2004,Barros2006,Liu2013}. Compared with atoms and linear molecules, CNTs combine a larger transverse extent with delocalized \(\pi\) electrons; together, these features can support a larger transition dipole moment for circumferential motion and thereby favor the optical injection of angular momentum. Their broader manifold of electronic excitations may also provide a wider range of optically accessible pumping channels. Previous studies have predicted that circularly polarized light can induce circumferential currents in CNTs and that CNTs can convert optical angular momentum into twisted excitonic states, supporting their description as fixed-axis angular-momentum converters~\cite{Kibis2021,Zang2022TwistedExcitons}. Strong-field photoemission studies and a recently demonstrated CNT-based electron source provide the complementary electron-release capability~\cite{Li2019ExtremeCNTPhotoemission,Guan2023CNTPhotoemission,Chen2025CNTSource}. CNTs therefore combine a structurally defined OAM axis, circumferential electronic motion, optical-helicity coupling, and electron emission within a single material platform. Moreover, their diameter, chirality, length, termination, doping, and surrounding environment provide a broad materials-design space that may be used to tune OAM preparation and electron release~\cite{Qian2020}.

\begin{figure*}[t]
  \centering
  \includegraphics[width=\textwidth]{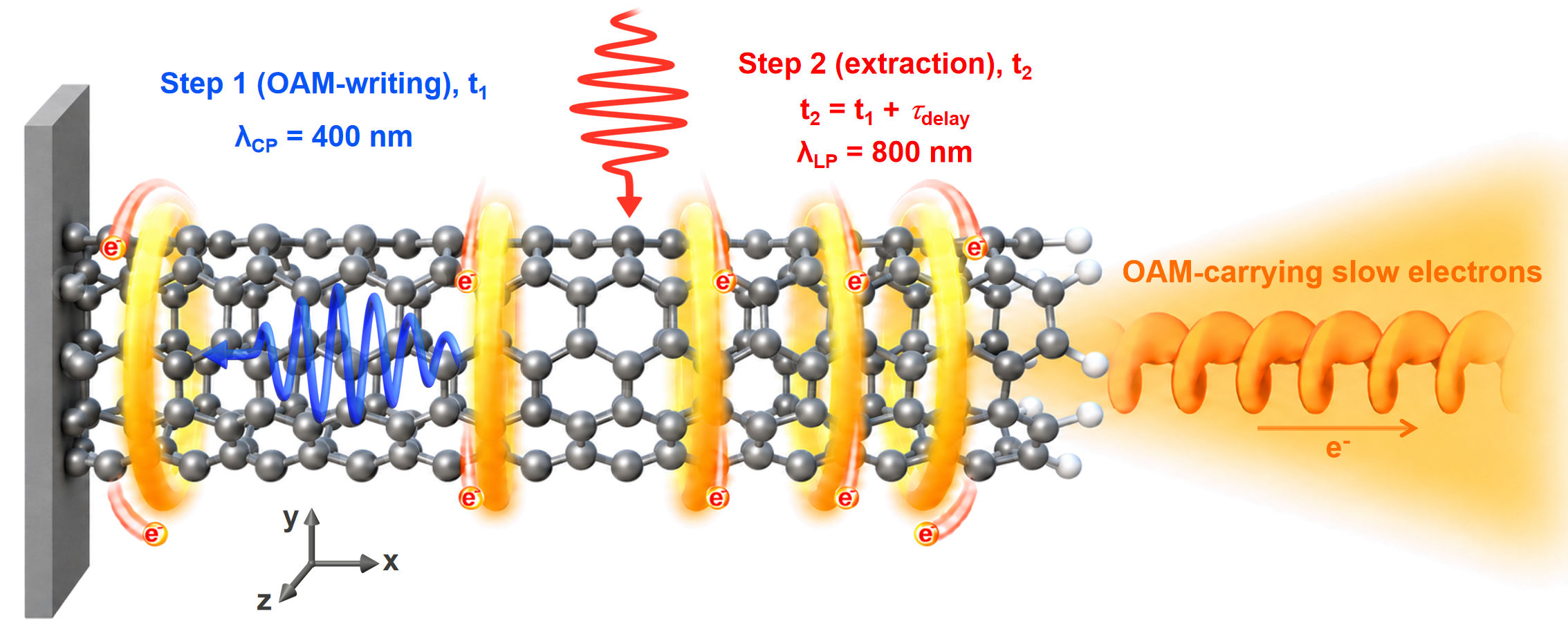}
  \caption{Schematic of the CNT electron source. \revision{Step 1:} A circularly polarized OAM-writing pulse couples to the transverse circumferential electronic motion of the CNT. \revision{Step 2:} A delayed extraction pulse, linearly polarized along the tube axis, releases the electrons prepared in \revision{Step 1}. Specific laser parameters are given in the Supplemental Material. The yellow--orange trajectories indicate the circumferential electron motion around the CNT. The emitted slow electrons carry \revision{axial OAM} about the CNT axis, as schematically represented by the orange helical structure. Carbon and hydrogen atoms are shown as gray and white spheres, respectively, and the coordinate system indicates the spatial orientation of the CNT.}
  \label{fig:schematic}
\end{figure*}

Using time-dependent density-functional theory (TDDFT) calculations, we show that hydrogen-terminated armchair CNTs can convert optical helicity into axial OAM carried by emitted slow electrons. Armchair CNTs, being nominally metallic or small-gap, facilitate low-energy optical pumping of circumferential electronic motion. Given the computational demands, we focus on two small-diameter armchair CNTs, CNT(3,3) and CNT(5,5). As shown in Fig.~\ref{fig:schematic}, a circularly polarized OAM-writing pulse drives helicity-dependent circumferential electronic motion; a delayed linearly polarized extraction pulse then releases \revision{slow electrons carrying axial OAM}. Unless otherwise specified, all circularly polarized pulses in this work are left-handed. We compare the CNT results with two molecular benchmarks under identical laser conditions: CO\(_2\), a compact linear molecule, and benzene, a cyclic molecule with a delocalized \(\pi\) system.

%% file: sections/methods.tex
All time-dependent electron dynamics were simulated using a modified version of the OCTOPUS 15.0 package~\cite{TancogneDejean2020Octopus}, which computes one-dimensional electron-density and axial-OAM distributions. We take the nanotube axis, which is also the OAM axis of interest, as the \emph{x} axis, with the \emph{yz} plane as the transverse plane. The electron-density and axial-OAM line distributions are denoted by \(\rho^{\mathrm{line}}(x,t)\) and \(L_{x}^{\mathrm{line}}(x,t)\), respectively. The differential line density is defined as \({\Delta\rho}^{\mathrm{line}}(x,t)=\rho^{\mathrm{line}}(x,t)-\rho^{\mathrm{line}}(x,0)\), where \(\rho^{\mathrm{line}}(x,0)\) is the ground-state line density. We integrate \(\rho^{\mathrm{line}}(x,t)\) and \(L_{x}^{\mathrm{line}}(x,t)\) over the integration region outside the target defined by the \emph{x}-space cuts shown in Figs.~\ref{fig:benchmark} and~\ref{fig:helicity}, yielding the collected charge \(n(t)\) and the collected axial OAM \(L_{\revision{x}}(t)\). We further define \(\widetilde{L}_{\revision{x}}(t)=L_{\revision{x}}(t)/n(t)\) as the mean axial OAM per ionized electron in the integration region; the signs of these OAM measures identify the handedness of the emitted axial OAM. Details of the CNT geometries, laser parameters, numerical settings, angular-momentum evaluation~\cite{Pickard2003GIPAW,Pickard2001MagneticResponse}, and integration-region definition are provided in the Supplemental Material.

%% file: sections/results.tex
\begin{figure*}[t]
  \centering
  \includegraphics[width=\textwidth]{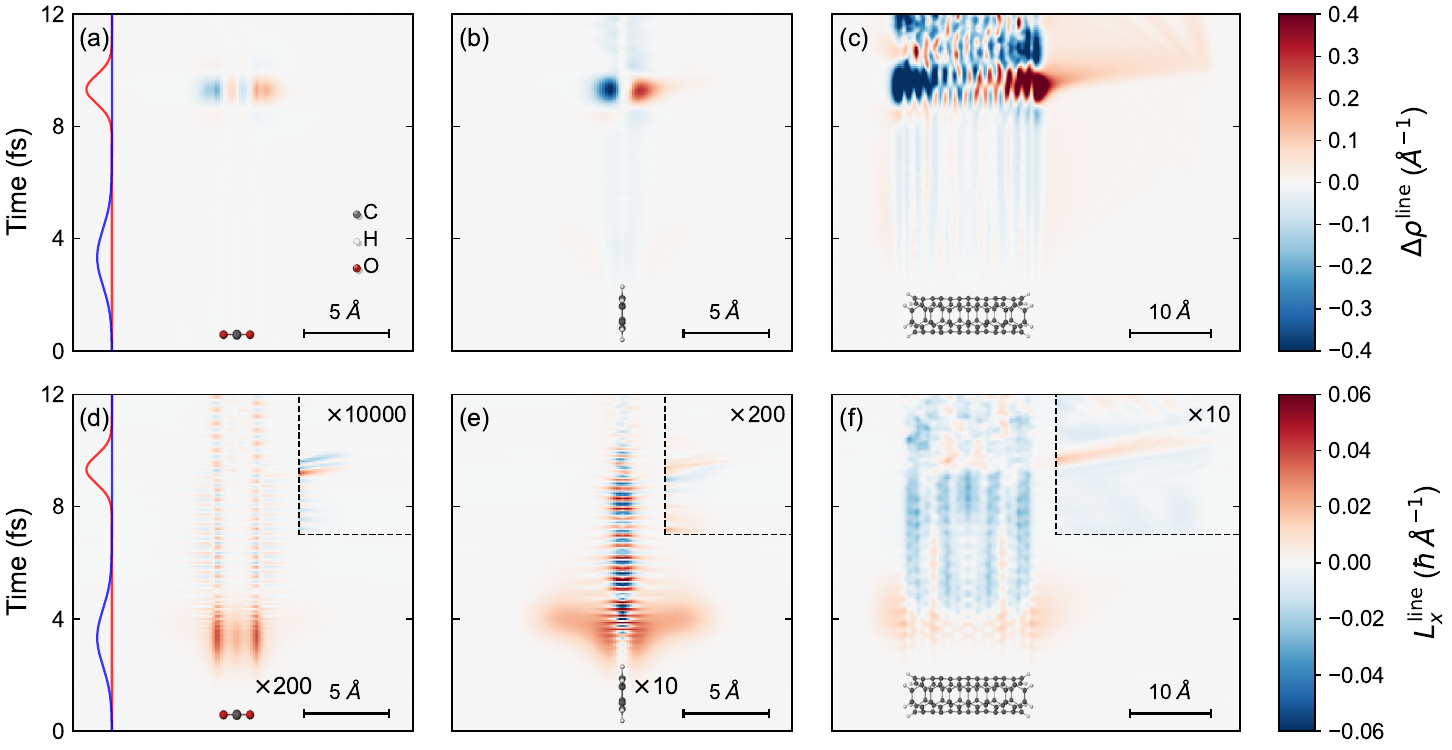}
  \caption{Time-dependent differential electron line density \({\Delta\rho}^{\mathrm{line}}(x,t)\) and axial-OAM line distribution \(L_{x}^{\mathrm{line}}(x,t)\) for CO\(_2\), benzene, and CNT(3,3). The horizontal axis is the x direction, with the positive direction pointing to the right, and the target is shown at its corresponding position in each panel. The leftmost curves depict the temporal profiles of the laser pulses. Panels (a--c) show \({\Delta\rho}^{\mathrm{line}}\); panels (d--f) show \(L_{x}^{\mathrm{line}}\). The global display factors for \(L_{x}^{\mathrm{line}}\) are \(\times\)200, \(\times\)10, and \(\times\)1 for CO\(_2\), benzene, and CNT(3,3), respectively. Dashed boxes mark the integration regions used to evaluate \(L_{\revision{x}}(t)\) and \(n(t)\). The signals within these boxes are displayed with the larger scaling factors indicated in the panels. Scale bars indicate 5 \AA{} for CO\(_2\) and benzene and 10 \AA{} for CNT(3,3). Color bars give \({\Delta\rho}^{\mathrm{line}}\) in \AA{}\textsuperscript{-1} and \(L_{x}^{\mathrm{line}}\) in \(\hbar\) \AA{}\textsuperscript{-1}.}
  \label{fig:benchmark}
\end{figure*}
\begin{figure}[t]
  \centering
  \includegraphics[width=\linewidth]{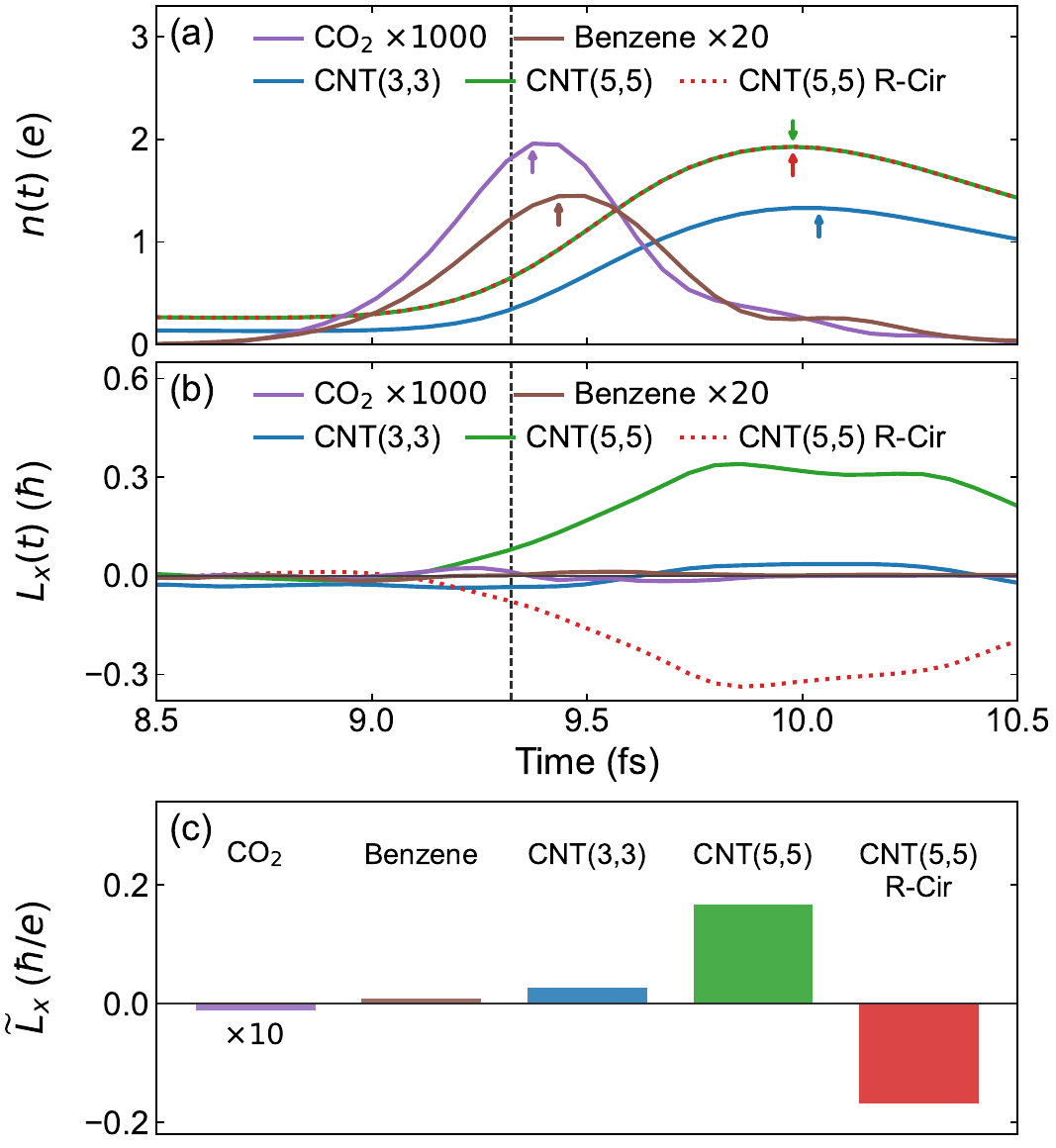}
  \caption{Integrated observables of the emitted wave packet. Panels (a) and (b) show the time-dependent collected charge \(n(t)\) and collected axial OAM \(L_{\revision{x}}(t)\), respectively, over the 8.5--10.5 fs interval. The vertical dashed line marks the peak of the linear extraction pulse. The CO\(_2\) and benzene traces are scaled by \(\times\)1000 and \(\times\)20, respectively. Panel (c) compares the mean axial OAM per ionized electron, \(\widetilde{L}_{\revision{x}}=L_{\revision{x}}/n\), evaluated at the respective peak-charge moments marked by the arrows in panel (a). For visibility in panel (c), the CO\(_2\) bar is scaled by a factor of 10; all other bars are unscaled. Positive and negative values indicate opposite handedness of the emitted axial OAM. ``R-Cir'' denotes a right-handed circularly polarized OAM-writing pulse.}
  \label{fig:integrated}
\end{figure}

Figure~\ref{fig:benchmark} compares the differential electron line density \({\Delta\rho}^{\mathrm{line}}(x,t)\) and axial-OAM line distribution \(L_{x}^{\mathrm{line}}(x,t)\) of CNT(3,3), the smaller nanotube considered here, with those of oriented CO\(_2\) and benzene under identical laser conditions. After the extraction pulse, CNT(3,3) produces \revision{a pronounced emitted electron wave packet carrying axial OAM}, whereas both molecular benchmarks show much weaker outgoing density and OAM signals; the CO\(_2\) and benzene OAM panels therefore require global display factors of \(\times\)200 and \(\times\)10, respectively. A free-electron estimate based on the propagation speed extracted from the emitted-density peak trajectory gives an axial kinetic-energy scale of about 8.5 eV, consistent with the slow-electron energy range (see Supplemental Material). The collected charge from CNT(3,3) reaches a maximum of \(1.33e\) (Fig.~\ref{fig:integrated}(a)), nearly three orders of magnitude larger than the corresponding maximum for CO\(_2\) and about 18 times larger than that for benzene. At the time of maximum collected charge for each system, the mean axial OAM per ionized electron \(\widetilde{L}_{\revision{x}}\) is \(0.027\,\hbar/e\) for CNT(3,3), while benzene gives a positive value about one third as large (Fig.~\ref{fig:integrated}(c)). The value for CO\(_2\) has the opposite sign but is not used in the quantitative comparison because the collected charge is very small. The collected charge also reaches its maximum slightly later for CNT(3,3) than for the two molecular systems (Fig.~\ref{fig:integrated}(a)). Similar delayed-emission behavior has been reported for laser-driven CNT electron sources~\cite{Chen2025CNTSource}.

\begin{figure}[t]
  \centering
  \includegraphics[width=\linewidth]{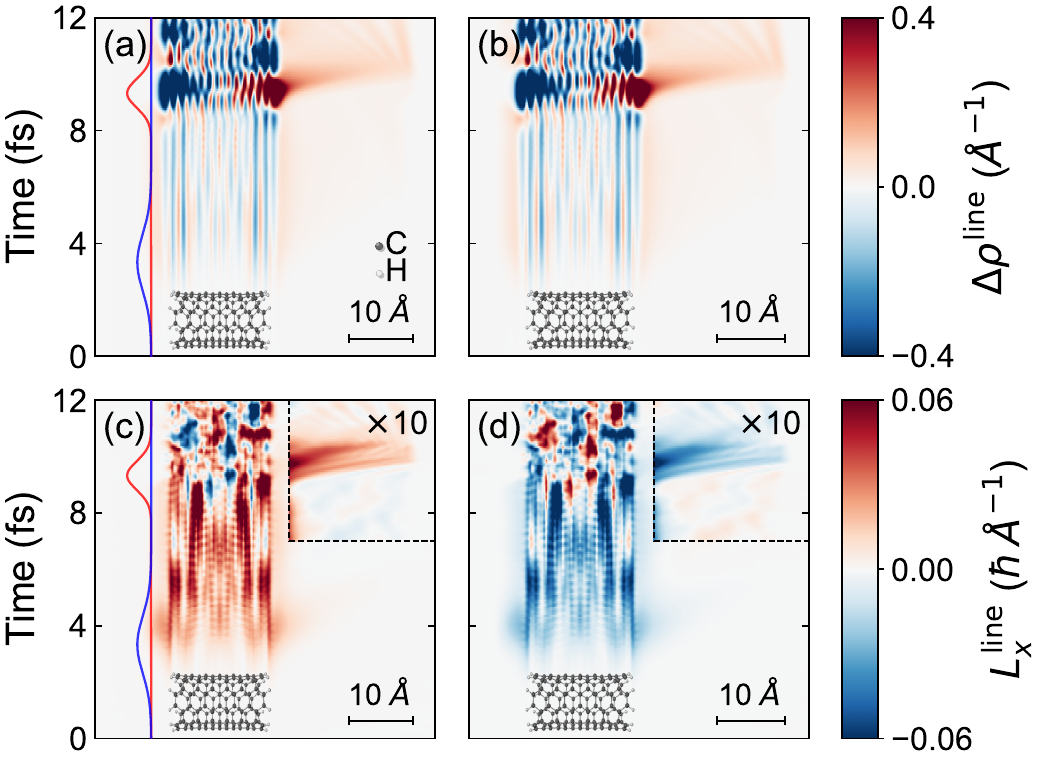}
  \caption{Time-dependent differential electron line density \({\Delta\rho}^{\mathrm{line}}(x,t)\) and axial-OAM line distribution \(L_{x}^{\mathrm{line}}(x,t)\) for CNT(5,5). Panels (a,b) show \({\Delta\rho}^{\mathrm{line}}\), and panels (c,d) show \(L_{x}^{\mathrm{line}}\). Panels (a,c) and (b,d) correspond to left- and right-handed circularly polarized OAM-writing pulses, respectively.}
  \label{fig:helicity}
\end{figure}

Increasing the tube diameter from CNT(3,3) to CNT(5,5), under otherwise identical laser conditions, produces a markedly stronger response (Figs.~\ref{fig:benchmark}--\ref{fig:integrated}). The maximum collected charge increases from \(1.33e\) to \(1.93e\), a 1.45-fold increase that is broadly comparable to the increase in system size from 78 to 130 carbon atoms (see Supplemental Material). The axial-OAM response grows much more strongly. At the time of maximum collected charge for each system, the collected axial OAM increases from \(0.036\,\hbar\) to \(0.322\,\hbar\), nearly ninefold, while the mean axial OAM per ionized electron \(\widetilde{L}_{\revision{x}}\) rises from \(0.027\,\hbar/e\) to \(0.167\,\hbar/e\), more than sixfold. The diameter dependence therefore cannot be attributed simply to the larger number of emitted electrons. A useful qualitative reference is provided by rotationally periodic ring systems~\cite{Peng2023SciAdvHHG}, in which neighboring quasi-angular-momentum channels obey the optical selection rule \(J-J'=\pm1\), and the corresponding transition-dipole matrix elements scale linearly with the ring radius, \(\langle \psi_J | x | \psi_{J+1} \rangle\sim R/2\). Although a CNT is not a simple ring, this radius scaling suggests why increasing the transverse extent can strengthen the coupling of the circular OAM-writing pulse to circumferential electronic motion. The CNT(3,3)--CNT(5,5) comparison therefore identifies tube diameter as a sensitive parameter for tuning the helicity-to-emitted-OAM conversion. The larger tube also shows a pronounced change in the spatial sign pattern of the emitted axial-OAM distribution. CNT(3,3) exhibits substantial positive and negative axial-OAM contributions at different positions (Fig.~\ref{fig:benchmark}(f)), which partially cancel upon spatial integration, whereas the CNT(5,5) distribution is positive at nearly all positions in the emitted wave packet (Fig.~\ref{fig:helicity}(c)). One possible origin of this difference is the higher rotational order of CNT(5,5), which may better suppress mixing between different axial-OAM components; establishing this connection quantitatively, however, will require a broader comparison across tube symmetries and diameters. The reduced cancellation between positive and negative contributions in CNT(5,5) results in \revision{a stronger predominance of positive axial OAM}.

We next test optical control of the OAM handedness in CNT(5,5). The CNT(5,5) calculation is repeated with the helicity of the circular OAM-writing pulse reversed, while all other laser conditions are kept unchanged (Figs.~\ref{fig:helicity} and~\ref{fig:integrated}). The differential-density pattern remains essentially unchanged, whereas the axial-OAM line distribution \(L_{x}^{\mathrm{line}}(x,t)\) reverses its sign both within the nanotube and in the emitted wave packet. Thus, reversing the OAM-writing pulse helicity reverses both the circumferential angular-momentum bias within the CNT and \revision{the sign of the emitted axial OAM}. The integrated observables quantify this control (Fig.~\ref{fig:integrated}). The two opposite optical helicities produce nearly identical maximum collected charges of \(1.93e\). At the common time of maximum collected charge, the mean axial OAM per ionized electron \(\widetilde{L}_{\revision{x}}\) flips from +\(0.167\,\hbar\)/\emph{e} to -\(0.168\,\hbar\)/\emph{e}. The nearly unchanged collected charge together with the sign-reversed OAM shows that the OAM-writing pulse helicity \revision{controls the sign of the emitted axial OAM} without substantially altering the electron yield.

%% file: sections/conclusion.tex
Taken together, these results provide a clear picture of \revision{emission of OAM-carrying slow electrons from CNTs} in terms of both performance and control. Under identical laser conditions, CNT(3,3) produces much larger collected charge and collected axial OAM than oriented CO\(_2\) and benzene. Increasing the tube diameter from CNT(3,3) to CNT(5,5) strengthens the OAM response far beyond the corresponding change in atomic count, identifying tube diameter as a materials-design parameter. Reversing the OAM-writing pulse helicity, in turn, reverses the sign of emitted axial OAM with little change in collected charge. The CNT converter demonstrated here turns \revision{the emission of OAM-carrying slow electrons} into a practical materials-and-laser design problem: the nanotube defines a fixed OAM axis, tube diameter tunes the conversion strength, and OAM-writing pulse helicity selects the handedness. Meanwhile, the broad materials design space (tube geometry, termination, and doping) and laser parameters (writing--extraction delay, intensity, duration, and frequency) provide coupled routes for further optimizing the helicity-to-emitted-OAM conversion. 

\revision{Beyond these emission properties, CNTs offer practical advantages for source integration. Slow electrons require a windowless vacuum path to the target or detector, where collisions with background gas can perturb their energy and OAM. A CNT mounted at a fixed position and orientation provides a stable laboratory-frame OAM axis, reducing the source-localization and molecular-alignment requirements associated with gas-phase emitters.} The scheme is compatible with existing CNT electron-source technology~\cite{Guan2023CNTPhotoemission,Chen2025CNTSource}. The large yield contrast relative to the molecular emitters considered here allows molecular targets to be placed near the source with reduced laser perturbation. This makes CNT emitters a promising platform for chirality-sensitive scattering with OAM-carrying slow electrons~\cite{Kolovertnova2026ChiralAsymmetry}. Such electrons provide a handedness channel complementary to that offered by spin-polarized slow electrons~\cite{Ray1999AsymmetricScattering,Dreiling2014VesterUlbricht}. They remain effective in chirality-sensitive scattering from molecules composed of light elements, where chiral selectivity mediated by spin--orbit interaction is weak.

%% file: sections/acknowledgments.tex
\begin{acknowledgments}
\revision{This work was supported by the National Key R\&D Program of China (2022YFA1604301, 2022YFA1602601), the National Natural Science Foundation of China (12374237, 12425411, 12434013, 12674425), the Natural Science Foundation of Jiangsu Province (BK20253027), and the Fundamental Research Funds for the Central Universities (WZJC202602002). We thank Prof.\ J.\ Chen and Prof.\ S.\ X.\ Tian for helpful discussions.}
\end{acknowledgments}